\documentclass[%
 reprint,
superscriptaddress,
 amsmath,amssymb,
 aps,
prl,
]{revtex4-2}

\usepackage{graphicx}
\usepackage{dcolumn}
\usepackage{bm}
\usepackage{hyperref}

\begin{document}

\preprint{APS/123-QED}

\title{Observation of Co-propagating Chiral Zero Modes in Magnetic Photonic Crystals}
\thanks{These authors contributed equally to this work.}%

\author{Zhongfu Li*}
 \affiliation{New Cornerstone Science Laboratory, Department of Physics, The University of Hong Kong, Hong Kong 999077, China}
 
 \author{Shaojie Ma*}
 \affiliation{Shanghai Engineering Research Centre of Ultra Precision Optical Manufacturing, Department of Optical Science and Engineering, School of Information Science and Technology, Fudan University, Shanghai 200433, China}
 
\author{Shuwei Li*}%
\affiliation{%
 National Engineering Research Center of Electromagnetic Radiation Control Materials, Key Laboratory of Multi-spectral Absorbing Materials and Structures of Ministry of Education, University of Electronic Science and Technology of China, Chengdu, China.
}%
\author{Oubo You}
\affiliation{New Cornerstone Science Laboratory, Department of Physics, The University of Hong Kong, Hong Kong 999077, China}

\author{Yachao Liu}
\affiliation{College of Electronics and Information Engineering, Shenzhen University,
518060 Shenzhen, China}

\author{Qingdong Yang}
\affiliation{New Cornerstone Science Laboratory, Department of Physics, The University of Hong Kong, Hong Kong 999077, China}

\author{Yuanjiang Xiang}
\affiliation{School of Physics and Electronics, Hunan University, Changsha, 410082, China}

\author{Peiheng Zhou}%
 \email{phzhou@uestc.edu.cn}
\affiliation{%
 National Engineering Research Center of Electromagnetic Radiation Control Materials, Key Laboratory of Multi-spectral Absorbing Materials and Structures of Ministry of Education, University of Electronic Science and Technology of China, Chengdu, China.
}%

\author{Shuang Zhang}
 \email{shuzhang@hku.hk}
\affiliation{New Cornerstone Science Laboratory, Department of Physics, The University of Hong Kong, Hong Kong 999077, China}

\begin{abstract}
Topological singularities, such as Weyl points and Dirac points, can give rise to unidirectional propagation channels known as chiral zero modes (CZMs) when subject to a magnetic field. These CZMs are responsible for intriguing phenomena like the chiral anomaly in quantum systems. The propagation direction of each CZM is determined by both the applied magnetic field and the topological charge of the singularity point. While counter-propagating CZMs have been observed in 2D and 3D systems, the realization of co-propagating CZMs has remained elusive. Here we present the first experimental observation of co-propagating CZMs in magnetic photonic crystals hosting a single pair of ideal Weyl points WPs. By manipulating the crystal's structural configuration, we spatially alter the locations of the WPs, creating pseudo-magnetic fields in opposite directions between them. This arrangement results in a pair of CZMs that possess the same group velocity and co-propagate. Our work opens up new possibilities for topological manipulation of wave propagation and may lead to advancements in optical waveguides, switches, and various other applications.
\end{abstract}

\maketitle


The emergence of topological materials or topological insulator \cite{wangReflectionFree2008,khanikaevPhotonic2013,hePhotonic2016,maGuiding2015,chengRobust2016,wuScheme2015,guoObservation2019,luSymmetryprotected2016,gaoTopologically2018,shalaevRobust2019a} offers a new and versatile platform for investigating unidirectional propagation of EM waves. A distinctive feature that emphasizes their uniqueness and counterintuitive nature within topological systems is the presence of topologically protected edge states or surface states\cite{wangReflectionFree2008,maGuiding2015,chengRobust2016,gaoTopologically2018,shalaevRobust2019a,ryuTopological2002,liFragile2020,guoThree2017,yangDirect2017,yangIdeal2018,chengVortical2020,maLinked2021,shaojiemaTopological2022,zhouObservation2020}. These topological states are immune to backscattering and disorder, making them ideal candidates for various applications in information processing and communications. One well-known example is the unidirectional edge states of quantum Hall systems1\cite{haldanePossible2008,klitzingNew1980,halperinQuantized1982,wangObservation2009,jinFloquet2022,luFloquet2021}, where the existence of magnetic field breaks time-reversal ($\mathcal{T}$) symmetry and results in quantized Hall conductance associated with an integer Chern number in Landau bands. Classical wave systems have exhibited similar behavior, supported by theoretical and experimental studies1\cite{wangObservation2009,skirloMultimode2014,skirloExperimental2015}. In addition to quantum Hall systems, extensive research has been conducted on edge states in quantum Valley Hall, quantum spin Hall systems\cite{wuScheme2015,liFragile2020,maAllSi2016,dongValley2017,luObservation2017,nohObservation2018}. These states are generally confined to the interface. Recent advancements have also revealed large-area topological waveguide states and one-way bulk states in heterostructures and two-dimensional photonics\cite{wangTopological2021a,wangValleylocked2020,chenPrediction2022}. 

While initial efforts have primarily focused on 2D systems, there has been significant interest in extending topological systems to higher dimensions. This includes exploring three-dimensional (3D) topological insulators as well as topological gapless states such as Weyl, Dirac, and nodal line systems. In particular, Weyl points (WPs) are linear crossing points of band structures in three-dimensional momentum space, acting as sources and sinks of Berry curvature fluxes. Weyl systems host a number of interesting phenomena, such as the presence of Fermi arc states at the interfaces and the formation of unidirectional propagation channels called chiral zero modes (CZMs) when subject to a strong magnetic field. CZMs are unique zeroth Landau levels that exhibit topologically protected bulk states and one-way propagating behavior. 

According to the Nielsen–Ninomiya no-go theorem, there exists equal number of Weyl degeneracies of opposite chiralities on a lattice\cite{nielsenAbsence1981}. Hence when a magnetic field is applied to a Weyl system, the CZMs formed by WPs of opposite chiralities would propagate along opposite directions. Recent developments in classical systems have demonstrated the existence of CZMs and Landau levels in Dirac and Weyl systems induced by a pseudo-magnetic field and the corresponding chiral one-way transport\cite{maGauge2023,jiaObservation2019,wenAcoustic2019,periAxialfieldinduced2019,grushinInhomogeneous2016,jiaExperimental2023}. However, in these systems, the preservation of $\mathcal{T}$ symmetry ensures that even in the vicinity of a certain $k$-point where a unidirectional chiral zero mode exists, there must also exist a corresponding mode propagating in the opposite direction at the -k counterpart, as shown in Fig. 1a. This implies that unidirectional transmission may be disrupted due to scattering between different $k$-modes. Hence it is highly desirable to have co-propagating CZMs such that the wave can remain unidirectional propagation in the presence of arbitrary scatterers, as illustrated by Fig. 1b. It has been theoretically proposed that co-propagating CZMs could be formed by applying both torsional deformation and external magnetic field\cite{pikulinChiral2016}. However, its implementation is very challenging and has not yet been realized experimentally. Here we present experimental demonstration of co-propagating CZMs in judiciously engineered photonic crystals with broken time reversal symmetry. Unidirectional propagation of electromagnetic waves (EM waves) has gained significant attention due to its potential applications in optical isolators, circulators, switches, and radio frequency communications\cite{wangReflectionFree2008,haldanePossible2008,sounasNonreciprocal2017,reiskarimianMagneticfree2016,liTunable2011,heTunable2010,yuComplete2009,serebryannikovOneway2009}. Our work may unlock novel approaches for achieving unidirectional transmission of bulk waves in topological materials. 

\begin{figure}
    \centering
    \includegraphics[width =\columnwidth]{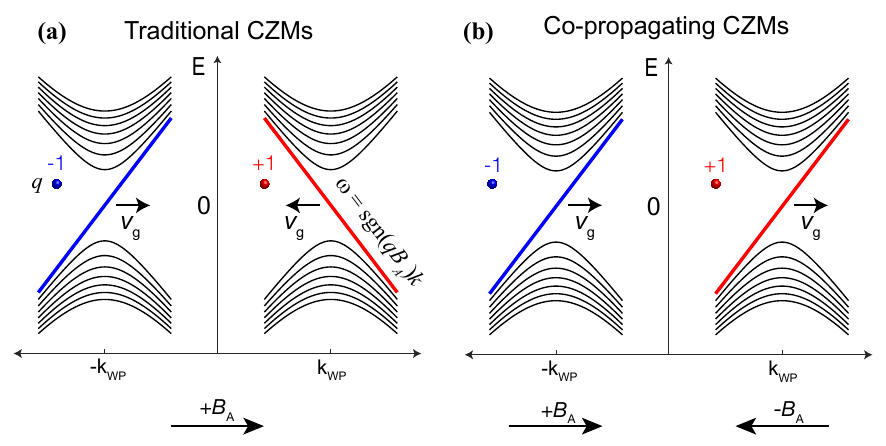}
    \caption{Dispersion of conventional and co-propagating CZMs. (a), The WPs under a magnetic field in the same direction generate two CZMs with opposite group velocities. The blue (red) dot indicates the -1 (+1) chirality (q) for WP. (b), The WPs under opposite pseudo-magnetic fields produce two CZMs with the same group velocity.}
    \label{fig1}
\end{figure}

The magnetic photonic crystal used in our work are composed of stacked copper with a staggered honeycomb hole lattice, where cylindrical yttrium iron garnet (YIG) rods are placed in the center of each honeycomb cell, as illustrated in Fig. 2A (additional details regarding sample fabrication can be found in the Methods section). When there is no bias magnetic field, the photonic crystal exhibits the properties of a 3D trivial insulator with a complete bandgap between 18.8 GHz and 19.3 GHz, as indicated by the blue-shaded region in Fig. 2b ($r_2=2.6mm$). However, when a non-zero magnetic field is applied along the YIG rods, near the K and K' points, the magnetic field exerts opposite effects on the bandgap due to the breaking of spatial inversion symmetry. It has been shown previously that only a pair of WPs could exist along the H'-K'-H direction in such a kind of photonic crystals induced by the bias magnetic field\cite{liuTopological2022}, as illustrated by the solid red line in Fig. 2b. Here we find that these WPs shift along the $k_z$ direction when the radius of the holes ($r_2$) is altered while maintaining the bias magnetic field constant. When considering an inhomogeneous photonic crystal with a gradually varying $r_2$, the k-space displacement can be regarded as a result of an equivalent vector potential $\vec{A}$. The artificial pseudo-magnetic field can be obtained from this vector potential as $\vec{B}=\nabla\times\vec{A}$. In this 3D magnetic photonic crystal, the WP with charge +1 (-1) shifts along the $+k_z$($-k_z$) direction and generates an artificial pseudo-magnetic field of $+B_A$ ($-B_A$) when the radius of the holes ($r_2$) gradually increases from 0.8 mm to 1.9 mm along the +y direction, as depicted in Fig. 2c. 

\begin{figure}
    \centering
    \includegraphics[width =\columnwidth]{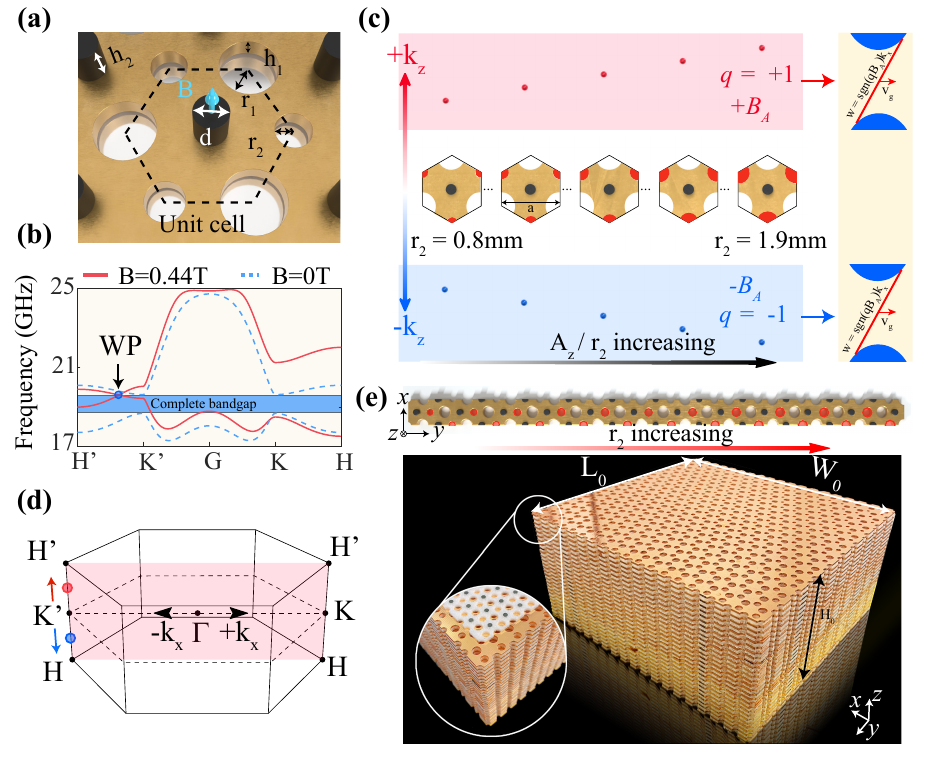}
    \caption{Photonic crystal design for achieving co-propagation CZMs. (a), Schematic of a gyromagnetic photonic crystal unit cell with a bias magnetic field. The crystal has a lattice constant (a) of 10 mm, composed of 1 mm thick (h1) copper with two kinds of holes (radius $r_1=2.2$mm and radius $r_1=1.4$mm) and YIG rods ($h_2=2$mm and $d=2.4$mm). (b), Simulated band structures without (dashed blue line) and with (solid red line) bias magnetic field. (c), Diagram illustrating the shifting of WPs with structural variations. The WPs with charge +1 (red dots) /-1 gradually shift along the $+k_z$/$-k_z$ direction and generate gauge fields ($+B_A$/$-B_A$). (d), Schematic of the Brillouin zone with one pair of WPs. The red cut plane is the $k_x-k_y$ plane. (e), Top panel shows a schematic of the supercell along the $y$ direction with red circles marking the varying holes. The bottom panel shows a photograph of the sample. The sample consists of 27 layers in the $z$ direction and 22- and 14-unit cells in the $x$- and $y$ directions, respectively.}
    \label{fig2}
\end{figure}
\begin{figure}
    \centering
    \includegraphics[width =\columnwidth]{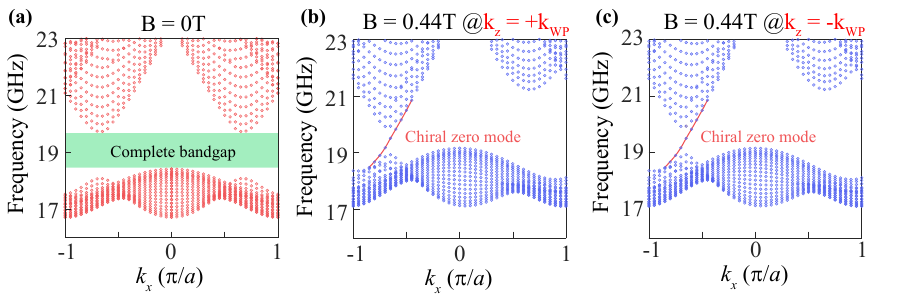}
    \caption{Simulation result of projection bands. (a), Simulated projection bands of supercell without bias magnetic field, exhibiting a complete band gap highlighted by the green shadow. (b), (c) Simulated projection bands of supercell with bias magnetic field (B = 0.44T) near two WPs ($k_{WP}=0.33\pi/h$). The solid red lines highlight the dispersion of CZMs ($n=0$).}
    \label{fig3}
\end{figure}
It is expected that the artificial pseudo-magnetic field generated by the shift of WPs would lead to formation of Landau levels. The minimal Hamiltonian of WPs can be expressed as $\mathcal{H}_{WP}=\sum_{i=x,y,z}{v_ik_i\sigma_i}$, where $\sigma_i$ represents the Pauli matrices and $v_i$ represents the Fermi velocities. According to the structure of the supercell shown in Fig. 2e, the gauge field can be written as $\vec{A}=\left(0,0,B_A\hat{y}\right)$, with a corresponding artificial pseudo-magnetic field $\vec{B}$ along the $x$ direction. Taking into account the coupling between the gauge field and the Hamiltonian of WPs, we arrive at the following equation:
\begin{equation}
 \mathcal{H}_{\mathrm{WP,LG}}=v_xk_x\sigma_x+v_y\left(-i\partial_y\right)\sigma_y+v_z\left(k_z-B_Ay\right)\sigma_z.   
\end{equation}

In this equation, we have replaced $k_y$ by $-i\partial_y$ since $k_y$ is ill-defined due to the $y$-dependent gauge field $A_z=B_A\hat{y}$. The dispersion of Landau levels can be obtained by solving a non-relativistic squared Hamiltonian (additional details can be found in the Sec. 1 of Supplementary Material):
\begin{equation}
    \mathcal{H}_{\mathrm{WP,LG}}^2=v_{x}^{2}v_{x}^{2}+v_{\bot}^{2}\left[\left|B_A\right|\left(2n+1\right)-B_A\mathrm{sgn}\left(v_x\right)q\sigma_x\right]
\end{equation}
where we have set $v_\bot=v_y=v_z$ for convenience and n is a non-negative integer. The last term $v_\bot^2B_A\mathrm{sgn}\left(v_x\right)q\sigma_x$ is the Zeeman term induced by the artificial pseudo-magnetic field. Previous works have demonstrated that each WP under the presence of an artificial pseudo-magnetic field exhibits symmetric relativistic high-order Landau levels and a single CZM\cite{maGauge2023,bellecObservation2020}. The dispersion of Landau levels can be expressed as: 
\begin{equation}
 \omega_{CZM}=\left\{\begin{matrix}\mathrm{sgn}\left(B_Aq\right)\left|v_x\right|k_x,&n=0\\\pm\sqrt{v_x^2k_x^2+v_{\bot}^2\left|B_A\right|\left(2n\pm1\right)},&n>0\\\end{matrix}\right.    
\end{equation}
where q is the topological charge of the WP. The sign of the chiral group velocity of the CZM depends only on the charge of the WP and the direction of the artificial pseudo-magnetic field. The corresponding dispersion of CZMs (red lines) and other Landau levels are shown in Fig. 2c. Interestingly, in our system, the group velocity directions sgn$\left(B_Aq\right)$ of the CZMs at two WPs are identical. This intriguing behavior can be attributed to the fact that the directions of the artificial pseudo-magnetic fields generated by WPs with opposite charges are also opposite, as depicted in Fig. 1b. When we apply the pseudo-magnetic field in this inhomogeneous photonic crystal, both WPs with $q=+1$ and $q=-1$ can generate CZMs with the same group velocity, i.e., co-propagating CZMs. This property enables our system to support unidirectional transmission of CZM, as depicted in Fig. 2d. 

Based on the band dispersion and the theoretical derivation of Landau level, we design an inhomogeneous photonic crystal to investigate the Landau levels and the co-propagating CZMs. The metamaterial consists of 20-unit cells, wherein the radius ($r_2$) exhibits spatial variation along the $y$ direction, ranging from 0.8 mm to 1.9 mm, marked by red circles in Fig. 2E. To confirm the existence of CZM and unidirectional transmission in our designed photonic crystal, we present simulation results in Fig. 3. First, we show the projection band without a bias magnetic field, in which the photonic crystal exhibits a complete bandgap ranging from 18.5 GHz to 20 GHz (Fig. 3a). When the bias magnetic field (B = 0.44 T) is applied, a CZM (solid red line in Fig. 3b and c) appears for both WPs located around $k_x=-2\pi/3a$, while a bandgap still exists along the $+k_x$ direction. The simulation results agree with our prediction of the constant direction of the CZM group velocity. These two co-propagating CZMs thus ensure that the topologically protected bulk states in our system only transport along a single direction.

To experimentally demonstrate the existence of co-propagating CZMs and the corresponding unidirectional transmission, we construct the designed sample (Fig. 2e) and employ near-field scanning techniques to probe the CZMs (more details about the setup and measurement of the experiment can be found in Methods). First, we measure the electric field distribution on the $x$-$z$ cut-plane at various frequencies by using a $z$-oriented dipole as the source located at the center of sample. The dispersion can be obtained by performing Fourier transformation of these electric field data along the $x$ direction. When the bias magnetic field equals zero, there is one complete bandgap between 18.5 GHz and 20 GHz (Fig. 4a). This bandgap is manifested in the transmission spectrum as a pronounced dip, depicted by the blue shadow in Fig. 4e. However, when the bias field is set to B = 0.44T, the formation of WPs along with the applied pseudo-magnetic field induces the co-propagating CZMs (Fig. 4b), which aligns well with the simulation. Figure 4d displays another CZM dispersion with the same group velocity for the other WP, confirming the presence of co-propagating CZMs in the system. Figure. 4f illustrates the transmission spectrum in the presence of bias magnetic field, which shows a significant non-reciprocal effect of about 25dB difference between the forward (S12) and backward (S21) transmission from 19.2GHz to 20.5GHz.
\begin{figure}
    \centering
    \includegraphics[width =\columnwidth]{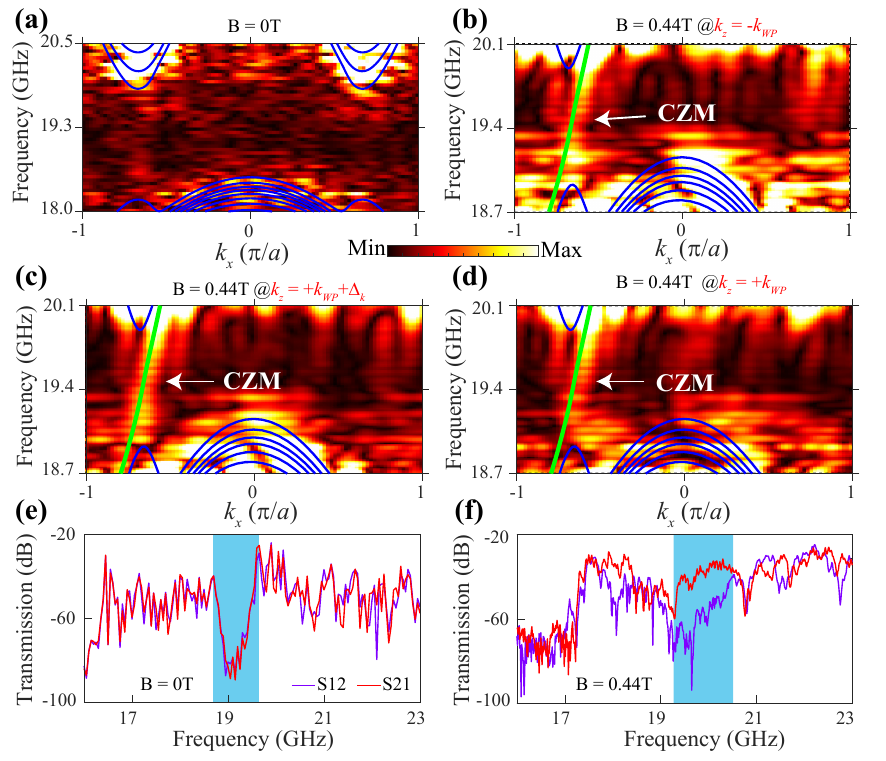}
    \caption{Experimental observation of CZMs. (a), Measured dispersion spectra without bias magnetic field. One complete band gap between 18.5 GHz and 20.0 GHz can be observed. (b)-(d), Dispersion spectra of CZM with bias magnetic field B = 0.44T at different kz. The blue and green line represent the bulk modes and CZMs, respectively. (e), (f), Transmission spectra without bias magnetic field and with the bias magnetic field. The blue shadow marks the frequency region of complete bandgap and unidirectional transmission.}
    \label{fig4}
\end{figure}
\begin{figure}
    \centering
    \includegraphics[width =\columnwidth]{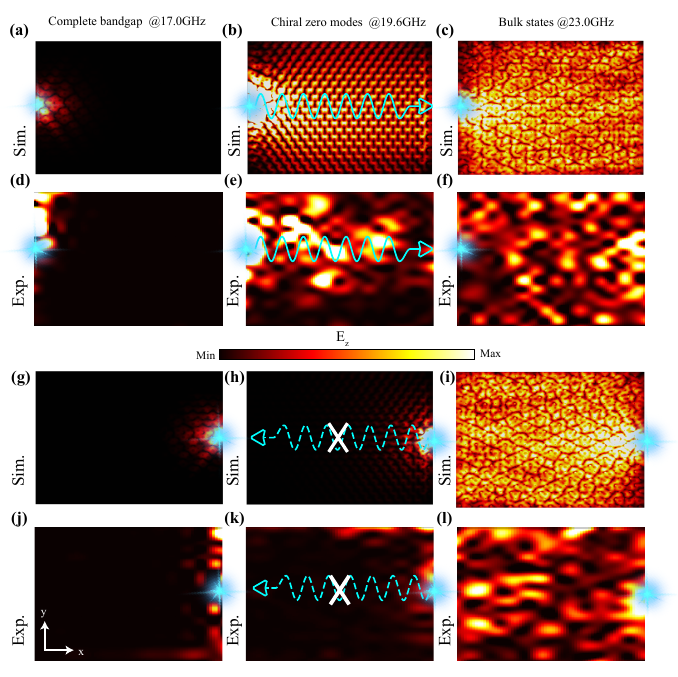}
    \caption{Simulated and measured unidirectional transmission excited by a z-oriented source at the $x-y$ plane with a height of $z=H_0/2$. Simulated [(a)-(c)] and measured [(d)-(f)] electric field distribution excited by a $z$-oriented source placed at the left edge. The sources are marked by the blue stars. The EM wave propagates in all directions for 23GHz and localized around the source for 17.0GHz, corresponding to bulk states and the lack of eigenstates within the complete gap. However, for 19.6 GHz, corresponding to the CZMs, the EM wave only propagate along the +$x$ direction. The experimental results match well with the simulation results. Simulated [(g)-(i)] and measured [(j)-(l)] electric field distribution excited by a $z$-oriented source placed at the right edge. At 19.6 GHz, the EM wave cannot propagate since there are no backward CZMs.}
    \label{fig5}
\end{figure}

Figure 5 displays the simulated (a-c and g-i) and measured (d-f and j-l) electric field distribution ($E_z$) in the $x-y$ cut-plane at three different frequencies: 17 GHz (complete bandgap), 19.6 GHz (CZMs) and 23.0 GHz (bulk states), under a bias magnetic field of B = 0.44 T (additional experimental results can be found in Sec. 4 of Supplementary Materials). In the simulation, periodic boundary conditions are applied in the $z$ direction, as well as scattering boundary conditions in other directions. At the CZM frequency, since both CZMs have a group velocity pointing in the +$x$ direction, EM waves can only be transmitted along the +$x$ direction, as shown in the full-wave simulation results (Fig. 5b). In the experiment, similar effect is observed. Specifically, when the dipole source is placed at the left edge of the sample, the EM wave can be transmitted to the right along the +$x$ direction (Fig. 5e). However, when the dipole source is positioned at the right edge of the sample, the EM wave is primarily confined to the source location, since there exists no mode with a group velocity pointing in the -$x$ direction (Fig. 5k). At frequency around 17 GHz (complete bandgap) and 23 GHz (bulk states), the EM waves respectively localize around the source (Fig. 5d and j) or transmit in all directions (Fig. 5f and l), regardless of the dipole source position. The above results match well with our theoretical simulation results, confirming the presence of co-propagating CZMs.

In conclusion, we have experimentally observed co-propagating CZMs, and the corresponding unidirectional transmission induced by pseudo-magnetic field in 3D magnetic photonic crystals. This is in sharp contrast to all previous studies that have demonstrated the existence of counter propagating CZMs. The co-propagating CZMs are capable of unidirectional transmission of bulk waves, immune from scattering by arbitrary scatterers and defects. Our work offers a novel approach to achieving unidirectional transmission of bulk waves in topological photonic crystals. In addition, our work fosters the exploration of CZMs in topological systems and paves the way for further development of the electromagnetic waves transportation.

\section*{ACKNOWLEDGMENTS}
This work was supported by the New Cornerstone Science Foundation, the Research Grants Council of Hong Kong (STG3/E-704/23-N, AoE/P-502/20 and 17309021). Work at University of Electronic Science and Technology of China was supported by the National Natural Science Foundation of China (grant no. 52022018 and 52021001).

The authors declare no conflict of interest.

\bibliography{co_propagating_CZM}

\end{document}